\documentstyle[preprint,aps]{revtex}

\input psfig.sty
\tightenlines 
\begin{document}
\draft
\title{Weak magnetism effects in the direct Urca processes in cooling
neutron stars}
\author{L. B. Leinson}
\address{Institute of Terrestrial Magnetism, Ionosphere and Radio Wave\\
Propagation RAS, 142190 Troitsk, Moscow Region, Russia\\
E-mail: leinson@izmiran.rssi.ru}
\maketitle

\begin{abstract}
In the mean field approximation, we study the effects of weak magnetism and
pseudoscalar interaction in the neutrino energy losses caused by the direct
Urca processes on relativistic nucleons in the degenerate baryon matter. Our
formula for the neutrino energy losses incorporates the effects of nucleon
recoil, parity violation, weak magnetism, and pseudoscalar interaction. For
numerical testing of our formula, we use a self-consistent relativistic
model of the multicomponent baryon matter. We found that, due to weak
magnetism effects, relativistic emissivities increase by approximately
40-50\%, while the pseudoscalar interaction only slightly suppresses the
energy losses, approximately by 5\%.
\end{abstract}


\widetext

The direct Urca processes, $n\rightarrow p+l+\bar{\nu}_{l}$, $\ \ \
p+l\rightarrow n+\nu _{l}$, where $l$ is either an electron or a muon, are
the most powerful neutrino reactions by which neutron stars lose their
energy. In spite of widely accepted importance of these reactions, the
relevant neutrino energy losses are not well investigated yet. By modern
scenarios, the central density of the star can be up to eight times larger
than the nuclear saturation density, what implies a substantially
relativistic motion of nucleons \cite{PBPELK97}. The relevant equation of
state of the matter is usually calculated in the relativistic approach \cite%
{LRP94}, while the energy losses are still calculated by the
non-relativistic formula obtained by Lattimer et al. \cite{Lat91} more than
ten years ago. Some aspects of this problem was studied by Leinson and P\'{e}%
rez \cite{PLB}, who have estimated relativistic effects of baryon recoil and
parity violation in the direct Urca processes. The above relativistic
effects increase substantially the neutrino emissivity in comparison with
the known non-relativistic prediction.

In the present Letter we derive a relativistic expression for neutrino
energy losses caused by direct Urca processes on nucleons by taking into
account also the effects of weak magnetism and pseudoscalar interaction. It
is known that weak magnetism plays an important role in core collapse
supernovae by increasing mean free paths of antineutrinos \cite{Bur00}, \cite%
{Hor02}. Our goal is to study the role of weak magnetism in radiation of
neutrinos and antineutrinos at the long-cooling epoch of neutron star.

We employ the Walecka-type relativistic model of baryon matter \cite{Serot},
where the baryons interact via exchange of $\sigma $, $\omega $, and $\rho $
mesons, and perform the calculation of the neutrino energy losses in the
mean field approximation. This approximation is widely used in the theory of
relativistic nuclear matter, and allows to calculate in a self-consistent
way the composition of the matter together with energies, and effective
masses of the baryons. The Lagrangian density, which includes the
interaction of a nucleon field $\Psi $ with a scalar field $\sigma $, a
vector field $\omega _{\mu }$ and an isovector field ${\bf b}_{\mu }$ of $%
\rho $-meson is of the form\footnote{%
In principle, the pion fields should be also included in the model. However,
the expectation value of the pion field equals zero, giving no contribution
to the mean fields. Therefore, only non-redundant terms are exhibited in the
Lagrangian density.}$^{,}$\footnote{%
In what follows we use the system of units $\hbar =c=1$ and the Boltzmann
constant $k_{B}=1$. Summation over repeated Greek indexes is assumed}%
\begin{eqnarray}
{\cal L} &=&\bar{\Psi}\left[ \gamma _{\mu }\left( i\partial ^{\mu
}-g_{\omega B}\omega ^{\mu }-\frac{1}{2}g_{\rho B}{\bf b}^{\mu }\cdot {\bf %
\tau }\right) -\left( M_{B}-g_{\sigma B}\sigma \right) \right] \Psi 
\nonumber \\
&&-\frac{1}{4}F_{\mu \nu }F^{\mu \nu }+\frac{1}{2}m_{\omega }^{2}\omega
_{\mu }\omega ^{\mu }-\frac{1}{4}{\bf B}_{\mu \nu }{\bf B}^{\mu \nu }+\frac{1%
}{2}m_{\rho }^{2}{\bf b}_{\mu }{\bf b}^{\mu }  \nonumber \\
&&+\frac{1}{2}\left( \partial _{\mu }\sigma \partial ^{\mu }\sigma
-m_{\sigma }^{2}\sigma ^{2}\right) -U\left( \sigma \right) +\bar{l}\left(
i\gamma _{\mu }\partial ^{\mu }-m_{l}\right) l,  \label{Lagr}
\end{eqnarray}%
Here $\Psi $ are the Dirac spinor fields for nucleons, ${\bf b}_{\mu }$ is
the isovector field of $\rho $-meson. We denote as ${\bf \tau }$ the isospin
operator, which acts on the nucleons of the bare mass $M$. The leptons are
represented only by electrons and muons, $l=e^{-},\mu ^{-}$, which are
included in the model as noninteracting particles. The field strength
tensors for the $\omega $ and $\rho $ mesons are $F_{\mu \nu }=\partial
_{\mu }\omega _{\nu }-\partial _{\nu }\omega _{\mu }$ and ${\bf B}_{\mu \nu
}=\partial _{\mu }{\bf b}_{\nu }-\partial _{\nu }{\bf b}_{\mu }$,
respectively. The potential $U\left( \sigma \right) $ represents the
self-interaction of the scalar field and is taken to be of the form 
\begin{equation}
U\left( \sigma \right) =\frac{1}{3}bM\left( g_{\sigma N}\sigma \right) ^{3}+%
\frac{1}{4}c\left( g_{\sigma N}\sigma \right) ^{4}.  \label{selfint}
\end{equation}%
In what follows we consider the mean field approximation widely used in the
theory of relativistic nuclear matter. In this approximation, the meson
fields are replaced with their expectation values%
\begin{equation}
\sigma \rightarrow \left\langle \sigma \right\rangle \equiv \sigma _{0},\,\
\ \omega ^{\mu }\rightarrow \left\langle \omega ^{\mu }\right\rangle \equiv
\omega _{0}\delta _{\mu 0},\,\ \ \ {\bf b}^{\mu }\rightarrow \left\langle 
{\bf b}^{\mu }\right\rangle \equiv \left( 0,0,\rho _{0}\right) \delta _{\mu
0}.  \label{avfield}
\end{equation}%
In this case only the nucleon fields must be quantized. This procedure
yields the following linear Dirac equation for the nucleon 
\begin{equation}
\left( i\partial _{\mu }\gamma ^{\mu }-g_{\omega }\gamma ^{0}\omega _{0}-%
\frac{1}{2}g_{\rho }\gamma ^{0}\rho _{0}\tau _{3}-\left( M-g_{\sigma }\sigma
_{0}\right) \right) \Psi \left( x\right) =0,  \label{DiracEq}
\end{equation}%
Here and below we denote as $\tau _{3}$, and $\tau _{\pm }=\left( \tau
_{1}\pm i\tau _{2}\right) /2$ the components of isospin operator, which act
on the isobaric doublet $\Psi \left( x\right) $ of nucleon field.

The stationary and uniform condensate fields equally shift the effective
masses 
\begin{equation}
M^{\ast }=M-g_{\sigma }\sigma _{0},  \label{Mef}
\end{equation}%
but lead to different potential energies of the proton and neutron 
\begin{equation}
U_{{\rm n}}=g_{\omega }\omega _{0}-\frac{1}{2}g_{\rho }\rho _{0},\,\ \ \ U_{%
{\rm p}}=g_{\omega }\omega _{0}+\frac{1}{2}g_{\rho }\rho _{0},  \label{Unp}
\end{equation}%
thus creating the energy gap $U_{{\rm n}}-U_{{\rm p}}=-g_{\rho }\rho _{0}$
between possible energies of protons and neutrons.

The exact solutions of Eq. (\ref{DiracEq}) can be found separately for
protons and for neutrons. In our case of a stationary and uniform system,
solutions are the spinor plane waves 
\begin{equation}
\psi _{{\rm n}}\left( x\right) =N_{{\rm n}}u_{{\rm n}}\left( P\right) \exp
\left( -iE_{{\rm n}}t+i{\bf pr}\right) ,  \label{nvf}
\end{equation}%
\begin{equation}
\psi _{{\rm p}}\left( x\right) =N_{{\rm p}}u_{{\rm p}}\left( P^{\prime
}\right) \exp \left( -iE_{{\rm p}}t+i{\bf p}^{\prime }{\bf r}\right) ,
\label{pvf}
\end{equation}%
where the neutron and proton energies are given by $E_{{\rm n}}\left( {\bf p}%
\right) =\sqrt{{\bf p}^{2}+M^{\ast 2}}+U_{{\rm n}}$, and $E_{{\rm p}}\left( 
{\bf p}^{\prime }\right) =\sqrt{{\bf p}^{\prime 2}+M^{\ast 2}}+U_{{\rm p}}$.
The free-like spinors $u_{{\rm n}}\left( P\right) $ and $u_{{\rm p}}\left(
P^{\prime }\right) $ are constructed from the kinetic momenta of the neutron
and the proton 
\begin{equation}
P^{\mu }=\left( E_{{\rm n}}-U_{{\rm n}},{\bf p}\right) =\left( \sqrt{{\bf p}%
^{2}+M^{\ast 2}},{\bf p}\right) ,  \label{kinN}
\end{equation}
\begin{equation}
P^{\prime \mu }=\left( E_{{\rm p}}-U_{{\rm p}},{\bf p}^{\prime }\right)
=\left( \sqrt{{\bf p}^{\prime 2}+M^{\ast 2}},{\bf p}^{\prime }\right) .
\label{kinP}
\end{equation}%
In what follows we denote by $\varepsilon =\sqrt{{\bf p}^{2}+M^{\ast 2}},\,\
\varepsilon ^{\prime }=\sqrt{{\bf p}^{\prime 2}+M^{\ast 2}}$ the kinetic
energy of the neutron and the proton respectively. So the normalization
factors are 
\begin{equation}
N_{{\rm n}}=\frac{1}{\sqrt{2\varepsilon }},\,\ \ \ \ \ N_{{\rm p}}=\frac{1}{%
\sqrt{2\varepsilon ^{\prime }}},  \label{norm}
\end{equation}%
and the single-particle energies can be written as $E_{{\rm n}}\left( {\bf p}%
\right) =\varepsilon +U_{{\rm n}}$, and $E_{{\rm p}}\left( {\bf p}^{\prime
}\right) =\varepsilon ^{\prime }+U_{{\rm p}}$.

We consider massless neutrinos of energy and momentum $k_{1}=\left( \omega
_{1},{\bf k}_{1}\right) $ with $\omega _{1}=\left| {\bf k}_{1}\right| . $
The energy-momentum of the final lepton $l=e^{-},\mu ^{-}$ of mass $m_{l} $
is denoted as $k_{2}=\left( \omega _{2},{\bf k}_{2}\right) $ with $\omega
_{2}=\sqrt{{\bf k}_{2}^{2}+m_{l}^{2}}$. In the lowest order in the Fermi
weak coupling constant $G_{F}$, the matrix element of the neutron beta decay
is of the form%
\begin{eqnarray}
\left\langle f\right| \left( S-1\right) \left| i\right\rangle &=&-i\frac{%
G_{F}C}{\sqrt{2}}N_{{\rm n}}N_{{\rm p}}\bar{u}_{l}\left( k_{2}\right) \gamma
_{\mu }\left( 1+\gamma _{5}\right) \nu \left( -k_{1}\right) \,_{{\rm p}%
}\left\langle P^{\prime }\right| J^{\mu }\left( 0\right) \left|
P\right\rangle _{{\rm n}}\times  \nonumber \\
&&\times \left( 2\pi \right) ^{4}\delta \left( E_{{\rm n}}-E_{{\rm p}%
}-\omega _{1}-\omega _{2}\right) \delta \left( {\bf p}-{\bf p}^{\prime }-%
{\bf k}_{1}-{\bf k}_{2}\right) ,  \label{S}
\end{eqnarray}%
were $C=\cos \theta _{C}=0\allowbreak .\,\allowbreak 973$ is the Cabibbo
factor. The effective charged weak current in the medium consists of the
polar vector and the axial vector, $J^{\mu }\left( x\right) =V^{\mu }\left(
x\right) +A^{\mu }\left( x\right) $.

Our goal now is to derive the nucleon matrix element of the charged weak
current in the medium. Consider first the polar-vector contribution. The
Lagrangian density (\ref{Lagr}) ensures a conserved isovector current \cite%
{SW97} 
\begin{equation}
{\bf T}^{\mu }=\frac{1}{2}\bar{\psi}\gamma ^{\mu }{\bf \tau }\psi +{\bf b}%
_{\nu }\times {\bf B}^{\nu \mu },\,\ \ \ \ \ \partial _{\mu }{\bf T}^{\mu
}=0.  \label{CIC}
\end{equation}%
Besides the directly nucleon contribution this current includes also the
contribution of the isovector field ${\bf b}^{\mu }$, which obeys the field
equations 
\begin{equation}
\partial _{\nu }{\bf B}^{\nu \mu }+m_{\rho }^{2}{\bf b}^{\mu }=\frac{1}{2}%
g_{\rho }\bar{\psi}\gamma ^{\mu }{\bf \tau }\psi ,\,\ \ \ \ \partial ^{\nu }%
{\bf b}_{\nu }=0.  \label{bEq}
\end{equation}%
By the use of Eq. (\ref{bEq}) the condition $\partial _{\mu }{\bf T}^{\mu
}=0 $ may be transformed as%
\begin{equation}
i\partial _{\mu }\left( \bar{\psi}\gamma ^{\mu }{\bf \tau }\psi \right)
=g_{\rho }{\bf b}_{\mu }\times \bar{\psi}\gamma ^{\mu }{\bf \tau }\psi .
\label{MFcic}
\end{equation}%
In the mean field approximation this gives 
\begin{equation}
i\partial _{\mu }\left( \bar{\psi}\gamma ^{\mu }\tau _{+}\psi \right)
=-g_{\rho }\rho _{0}\bar{\psi}\gamma ^{0}\tau _{+}\psi .  \label{ncnc}
\end{equation}%
By introducing the covariant derivative 
\begin{equation}
D_{\mu }=\left( \frac{\partial }{\partial t}-ig_{\rho }\rho _{0},\,{\bf %
\nabla }\right) ,  \label{cd}
\end{equation}%
we can recast the Eq. (\ref{ncnc}) to the following form%
\begin{equation}
D_{\mu }\left( \bar{\psi}\gamma ^{\mu }\tau _{+}\psi \right) =0.  \label{cc}
\end{equation}%
At the level of matrix elements this can be written as%
\begin{equation}
\bar{u}_{{\rm p}}\left( P^{\prime }\right) q_{\mu }\gamma ^{\mu }u_{{\rm n}%
}\left( P\right) =0,  \label{orthq}
\end{equation}%
where $q_{\mu }$ is the kinetic momentum transfer 
\begin{equation}
q^{\mu }=\left( E_{{\rm n}}-E_{{\rm p}}+g_{\rho }\rho _{0},\,{\bf p-p}%
^{\prime }\right) =\left( \varepsilon -\varepsilon ^{\prime },{\bf p-p}%
^{\prime }\right)  \label{q}
\end{equation}%
Thus the matrix element of the transition current is orthogonal to the {\em %
kinetic} momentum transfer, but not to the total momentum transfer from the
nucleon\footnote{%
Some consequences of this fact for the weak response functions of the medium
are discussed in \cite{PLB}}. Note that this effect originates not from a
special form (\ref{CIC}) of the conserved isovector current in the medium
but is caused by the energy gap between the proton and neutron spectrums.
This follows directly from the Dirac equation (\ref{DiracEq}), which ensures
the Eq. (\ref{orthq}) with 
\begin{equation}
q^{\mu }=P^{\mu }-P^{\prime \mu }=\left( E_{{\rm n}}-E_{{\rm p}}-U_{{\rm n}%
}+U_{{\rm p}},\,{\bf p-p}^{\prime }\right) ,  \label{qq}
\end{equation}%
which coincides with Eq. (\ref{q}) because $U_{{\rm n}}-U_{{\rm p}}=-g_{\rho
}\rho _{0}$.

By the use of the isovector current (\ref{CIC}) one can construct the
conserved electromagnetic current in the medium%
\begin{equation}
J_{{\rm em}}^{\mu }=\frac{1}{2}\bar{\psi}\gamma ^{\mu }\psi +T_{3}^{\mu }+%
\frac{1}{2M}\,\partial _{\nu }\left( \bar{\Psi}\lambda \sigma ^{\mu \nu
}\Psi \right) ,\,\ \ \ \ \ \ \partial _{\mu }J_{{\rm em}}^{\mu }=0.
\label{Jem}
\end{equation}%
The last term in\ Eq. (\ref{Jem}) is the Pauli contribution, where $2\sigma
^{\mu \nu }=\gamma ^{\mu }\gamma ^{\nu }-\gamma ^{\nu }\gamma ^{\mu }$ and 
\begin{equation}
\lambda =\lambda _{{\rm p}}\frac{1}{2}\left( 1+\tau _{3}\right) +\lambda _{%
{\rm n}}\frac{1}{2}\left( 1-\tau _{3}\right) .  \label{lam}
\end{equation}%
In the mean field approximation, we replace the magnetic formfactors of the
nucleon with anomalous magnetic moments of the proton and the neutron, $%
\lambda _{{\rm p}}=1.7928$ and $\lambda _{{\rm n}}=-1.9132$.

By the conserved-vector-current theory (CVC), the nucleon matrix element of
the charged vector weak current is given by 
\begin{equation}
_{{\rm p}}{}\left\langle P^{\prime }\right| V^{\mu }\left| P\right\rangle _{%
{\rm n}}=\,_{{\rm p{}}}\left\langle P^{\prime }\right| J_{{\rm em}}^{\mu
}\left| P\right\rangle _{{\rm p}}-\,_{{\rm n}}{}\left\langle P^{\prime
}\right| J_{{\rm em}}^{\mu }\left| P\right\rangle _{{\rm n}}.  \label{VJ}
\end{equation}
Thus, in the mean field approximation, we obtain 
\begin{equation}
_{{\rm p}}{}\!\left\langle P^{\prime }\right| V^{\mu }\left( 0\right) \left|
P\right\rangle _{{\rm n}}=\bar{u}_{{\rm p}}\left( P^{\prime }\right) \left[
\gamma ^{\mu }+\frac{\lambda _{{\rm p}}-\lambda _{{\rm n}}}{2M}\sigma ^{\mu
\nu }q_{\nu }\right] u_{{\rm n}}\left( P\right) .  \label{wcv}
\end{equation}%
The second term in Eq. (\ref{wcv}), describes the weak magnetism effects. By
the use of Eq. (\ref{DiracEq}) we find 
\begin{equation}
_{{\rm p}}{}\!\left\langle P^{\prime }\right| q_{\mu }V^{\mu }\left(
0\right) \left| P\right\rangle _{{\rm n}}=0.  \label{qV}
\end{equation}

Consider now the axial-vector charged current. This current is responsible
for both the $np$ transitions and the pion decay. In the limit of chiral
symmetry, $m_{\pi }\rightarrow 0$, the axial-vector current must be
conserved. In the medium with $\rho $ meson condensate, this implies 
\begin{equation}
\lim_{m_{\pi }\rightarrow 0}D_{\mu }A^{\mu }\left( x\right) =0,  \label{DA}
\end{equation}%
where the covariant derivative $D_{\mu }$ is defined by Eq. (\ref{cd}). At
the finite mass of a pion, $m_{\pi }$, the axial-vector charged current is
connected to the field $\pi _{-}\left( x\right) =\left( \pi _{1}+i\pi
_{2}\right) /\sqrt{2}$ of $\pi ^{-}$ meson. For a free space, this relation
is known as the hypothesis of partial conservation of the axial current
(PCAC). In the medium the PCAC takes the form 
\begin{equation}
D_{\mu }A^{\mu }\left( x\right) =m_{\pi }^{2}f_{\pi }\pi _{-}\left( x\right)
,  \label{PCAC}
\end{equation}%
where $m_{\pi }=139\,MeV$ is the mass of $\pi $-meson, and $f_{\pi }$ is the
pion decay constant.

With allowing for interactions of the pions with nucleons and $\rho $ mesons
the Lagrangian density for the pion field is of the form \cite{SW97}%
\begin{equation}
{\cal L}_{\pi }=\frac{1}{2}\left[ \left( \partial _{\mu }{\bf \pi }-g_{\rho }%
{\bf b}_{\mu }\times {\bf \pi }\right) \cdot \left( \partial ^{\mu }{\bf \pi 
}-g_{\rho }{\bf b}^{\mu }\times {\bf \pi }\right) -m_{\pi }^{2}{\bf \pi
\cdot \pi }\right] +ig_{\pi }\bar{\psi}\gamma _{5}{\bf \tau \cdot \pi }\psi 
\text{.}  \label{Lpi}
\end{equation}%
In the mean field approximation this results in the following equation for
the field of $\pi ^{-}$ meson%
\begin{equation}
\left( \left( i\partial ^{0}+g_{\rho }\rho _{0}\right) ^{2}-\left( i\nabla
\right) ^{2}-m_{\pi }^{2}\right) \pi _{-}\left( x\right) =-\sqrt{2}ig_{\pi }%
\bar{\psi}_{{\rm p}}\gamma _{5}\psi _{{\rm n}},  \label{pifield}
\end{equation}%
where $g_{\pi }$ is the pion-nucleon coupling constant.

For the nucleon transition of our interest, the Eq. (\ref{PCAC}) gives 
\begin{equation}
\,_{{\rm p}}\!\left\langle P^{\prime }\right| q_{\mu }A^{\mu }\left(
0\right) \left| P\right\rangle _{{\rm n}}=im_{\pi }^{2}f_{\pi }\,_{{\rm p}%
}\left\langle P^{\prime }\right| \pi _{-}\left( 0\right) \left|
P\right\rangle _{{\rm n}}  \label{npPCAC}
\end{equation}%
Here the right-hand side can be calculated by the use of Eq. (\ref{pifield}%
). We obtain%
\begin{equation}
\,_{{\rm p}}\!\left\langle P^{\prime }\right| q_{\mu }A^{\mu }\left(
0\right) \left| P\right\rangle _{{\rm n}}=-\frac{\sqrt{2}m_{\pi }^{2}f_{\pi
}g_{\pi }}{m_{\pi }^{2}-q^{2}}\bar{u}_{{\rm p}}\left( P^{\prime }\right)
\gamma _{5}u_{{\rm n}}\left( P\right) \text{.}  \label{mePCAC}
\end{equation}%
This equation allows to derive the nucleon matrix element of the
axial-vector charged current. Really, to construct the axial-vector matrix
element of the charged current, caused by the nucleon transition, we have
only two independent pseudovectors, consistent with invariance of strong
interactions under $T_{2}$ isospin transformation, namely: $\bar{u}_{{\rm p}%
}\left( P^{\prime }\right) \gamma ^{\mu }\gamma _{5}u_{{\rm n}}\left(
P\right) $, and $\bar{u}_{{\rm p}}\left( P^{\prime }\right) q^{\mu }\gamma
_{5}u_{{\rm n}}\left( P\right) $. This means that the matrix element of the
axial-vector charged current is of following general form 
\begin{equation}
_{{\rm p}}\!\left\langle P^{\prime }\right| A^{\mu }\left( 0\right) \left|
P\right\rangle _{{\rm n}}=C_{A}\,\bar{u}_{{\rm p}}\left( P^{\prime }\right)
\left( \gamma ^{\mu }\gamma _{5}+F_{q}\,q^{\mu }\gamma _{5}\right) u_{{\rm n}%
}\left( P\right) .  \label{AmeG}
\end{equation}%
Here, in the mean field approximation, we set $C_{A}\simeq 1.26$, while $%
F_{q}$ is the form-factor to be chosen to satisfy the Eq. (\ref{mePCAC}),
which now reads%
\begin{equation}
C_{A}\,\left( -2M^{\ast }+F_{q}\,q^{2}\right) \bar{u}_{{\rm p}}\left(
P^{\prime }\right) \gamma _{5}u_{{\rm n}}\left( P\right) =-\frac{\sqrt{2}%
m_{\pi }^{2}f_{\pi }g_{\pi }}{m_{\pi }^{2}-q^{2}}\bar{u}_{{\rm p}}\left(
P^{\prime }\right) \gamma _{5}u_{{\rm n}}\left( P\right) .  \label{lhs}
\end{equation}
Thus%
\begin{equation}
C_{A}\,\left( 2M^{\ast }-F_{q}\,q^{2}\right) =\frac{\sqrt{2}m_{\pi
}^{2}f_{\pi }g_{\pi }}{m_{\pi }^{2}-q^{2}}.  \label{Feq}
\end{equation}%
In the mean field approximation, we assume that the coupling constants are
independent of the momentum transfer. By setting $q^{2}=0$ in Eq. (\ref{Feq}%
) we obtain the Goldberger - Treiman relation $f_{\pi }g_{\pi }=\sqrt{2}%
M^{\ast }C_{A}$. By inserting this in (\ref{Feq}) we find%
\begin{equation}
F_{q}\,=-\,\frac{2M^{\ast }}{\left( m_{\pi }^{2}-q^{2}\right) }.  \label{Fq}
\end{equation}%
Thus, with taking into account Eqs. (\ref{wcv}), (\ref{AmeG}), and (\ref{Fq}%
), the total matrix element of the neutron beta decay is found to be%
\begin{eqnarray}
{\cal M}_{fi} &=&-i\frac{G_{F}C}{\sqrt{2}}\bar{u}_{l}\left( k_{2}\right)
\gamma _{\mu }\left( 1+\gamma _{5}\right) \nu \left( -k_{1}\right) \,\times
\label{mend} \\
&&\times \bar{u}_{{\rm p}}\left( P^{\prime }\right) \left[ C_{V}\gamma ^{\mu
}+\frac{1}{2M}C_{M}\sigma ^{\mu \nu }q_{\nu }+C_{A}\left( \gamma ^{\mu
}\gamma _{5}+F_{q}\,q^{\mu }\gamma _{5}\right) \right] u_{{\rm n}}\left(
P\right) ,  \nonumber
\end{eqnarray}%
where, in the mean field approximation, we assume 
\begin{equation}
C_{V}=1,\,\ \ \ \ \ C_{M}=\lambda _{p}-\lambda _{n}\simeq 3.7,\,\ \ \
C_{A}=1.26.  \label{ff}
\end{equation}%
Note that the matrix element obtained is of the same form as that for the
neutron decay in a free space, but with the total momentum transfer replaced
with the kinetic momentum transfer. Due to the difference in the neutron and
proton potential energy, the kinetic momentum transfer 
\begin{equation}
q=P-P^{\prime }=\left( \varepsilon -\varepsilon ^{\prime },{\bf p-p}^{\prime
}\right)  \label{kintr}
\end{equation}%
to be used in the matrix element (\ref{mend}) differs from the total
momentum of the final lepton pair 
\begin{equation}
K\equiv k_{1}+k_{2}=\left( \varepsilon -\varepsilon ^{\prime }+U_{{\rm n}%
}-U_{{\rm p}},\,{\bf p-p}^{\prime }\right)  \label{Q}
\end{equation}%
This ensures $K^{2}>0$, while $q^{2}=\left( \varepsilon -\varepsilon
^{\prime }\right) ^{2}-\left( {\bf p}-{\bf p}^{\prime }\right) ^{2}<0$ .

The square of the matrix element of the reaction summed over spins of
initial and final particles is found to be\footnote{%
Here we rectify an error made in the journal version of the paper. The
correct expression can be obtained from that published in \cite{PLB02}, \cite%
{NPA02} by simple replacement $C_{M}\rightarrow C_{M}/2$. The author is
grateful to M. Prakash and S. Ratkovi\'{c} who have pointed out this error.}%
: 
\begin{eqnarray}
\left| {\cal M}_{fi}\right| ^{2} &=&32G_{F}^{2}C^{2}\left[ \left(
C_{A}^{2}-C_{V}^{2}\right) M^{\ast 2}\left( k_{1}k_{2}\right) +\left(
C_{A}-C_{V}\right) ^{2}\left( k_{1}P_{2}\right) \left( k_{2}P_{1}\right)
\right.  \nonumber \\
&&+\left( C_{A}+C_{V}\right) ^{2}\left( k_{1}P_{1}\right) \left(
k_{2}P_{2}\right)  \nonumber \\
&&+C_{M}\frac{M^{\ast }}{M}\left[ 2C_{A}\left( \left( k_{1}P_{1}\right)
\left( k_{2}P_{2}\right) -\left( k_{1}P_{2}\right) \left( k_{2}P_{1}\right)
\right) \right.  \nonumber \\
&&\left. +C_{V}\left( \left( k_{1}k_{2}\right) \left( P_{1}P_{2}-M^{\ast
2}\right) -\left( k_{1}P_{1}-k_{1}P_{2}\right) \left(
k_{2}P_{1}-k_{2}P_{2}\right) \right) \right]  \nonumber \\
&&-\frac{C_{M}^{2}}{4M^{2}}\left[ M^{\ast 2}\left( k_{1}P_{2}\right) \left(
3\left( k_{2}P_{2}\right) -\left( k_{2}P_{1}\right) \right) \right. 
\nonumber \\
&&+M^{\ast 2}\left( k_{1}P_{1}\right) \left( 3\left( k_{2}P_{1}\right)
-\left( k_{2}P_{2}\right) \right) +\left( k_{1}k_{2}\right) \left(
P_{1}P_{2}-M^{\ast 2}\right) ^{2}  \nonumber \\
&&\left. -\left( k_{1}P_{1}+k_{1}P_{2}\right) \left(
k_{2}P_{1}+k_{2}P_{2}\right) \left( P_{1}P_{2}\right) \right]  \nonumber \\
&&+C_{A}^{2}F_{q}\left( 2M^{\ast }+F_{q}\left( M^{\ast 2}-\left(
P_{1}P_{2}\right) \right) \right) \left[ \left( k_{1}k_{2}\right) \left(
M^{\ast 2}-\left( P_{1}P_{2}\right) \right) \right.  \nonumber \\
&&\left. \left. -\left( k_{1}P_{1}-k_{1}P_{2}\right) \left(
k_{2}P_{1}-k_{2}P_{2}\right) \right] \right]  \label{MatrEl}
\end{eqnarray}%
with $P_{1}=\left( \varepsilon ,{\bf p}\right) $ and $P_{2}=\left(
\varepsilon ^{\prime },{\bf p}^{\prime }\right) $.

We consider the total energy which is emitted into neutrino and antineutrino
per unit volume and time. Within beta equilibrium, the inverse reaction $%
p+l\rightarrow n+\nu _{l}$ corresponding to a capture of the lepton $l$,
gives the same emissivity as the beta decay, but in neutrinos. Thus, the
total energy loss $Q$ for the Urca processes is twice more than that caused
by the beta decay. Taking this into account by Fermi's ''golden'' rule we
have%
\begin{eqnarray}
Q &=&2\int \frac{d^{3}k_{2}d^{3}k_{1}d^{3}pd^{3}p^{\prime }}{(2\pi
)^{12}2\omega _{2}2\omega _{1}2\varepsilon 2\varepsilon ^{\prime }}\left| 
{\cal M}_{fi}\right| ^{2}\omega _{1}\,f_{{\rm n}}\left( 1-f_{{\rm p}}\right)
\left( 1-f_{l}\right)  \nonumber \\
&&\times \left( 2\pi \right) ^{4}\delta \left( E_{{\rm n}}\left( {\bf p}%
\right) -E_{{\rm p}}\left( {\bf p}^{\prime }\right) -\omega _{1}-\omega
_{2}\right) \delta \left( {\bf p}-{\bf p}^{\prime }-{\bf k}_{1}-{\bf k}%
_{2}\right) .  \label{gold}
\end{eqnarray}%
Antineutrinos are assumed to be freely escaping. The distribution function
of initial neutrons as well as blocking of final states of the proton and
the lepton $l$ are taken into account by the Pauli blocking-factor $\,f_{%
{\rm n}}\left( 1-f_{{\rm p}}\right) \left( 1-f_{l}\right) $. The Fermi-Dirac
distribution function of leptons is given by 
\begin{equation}
f_{l}\left( \omega _{2}\right) =\frac{1}{\exp \left( \omega _{2}-\mu
_{l}\right) /T+1},  \label{fl}
\end{equation}%
while the individual Fermi distributions of nucleons are of the form 
\begin{equation}
f_{{\rm n}}\left( \varepsilon \right) =\frac{1}{\exp \left( \left(
\varepsilon +U_{{\rm n}}-\mu _{{\rm n}}\right) /T\right) +1},  \label{fn}
\end{equation}%
\begin{equation}
f_{{\rm p}}\left( \varepsilon ^{\prime }\right) =\frac{1}{\exp \left( \left(
\varepsilon ^{\prime }+U_{{\rm p}}-\mu _{{\rm p}}\right) /T\right) +1},
\label{fp}
\end{equation}

By neglecting the chemical potential of escaping neutrinos, we can write the
condition of chemical equilibrium as $\mu _{l}=\mu _{{\rm n}}-\mu _{{\rm p}}$%
. Then by the use of the energy conservation equation, $\varepsilon +U_{{\rm %
n}}=\varepsilon ^{\prime }+U_{{\rm p}}+\omega _{2}+\omega _{1}$, and taking
the total energy of the final lepton and antineutrino as $\omega _{2}+\omega
_{1}=\mu _{l}+\omega ^{\prime }$ we can recast the blocking-factor as 
\begin{eqnarray}
&&\,f_{{\rm n}}\left( \varepsilon \right) \left( 1-f_{{\rm p}}\left(
\varepsilon ^{\prime }\right) \right) \left( 1-f_{l}\left( \omega
_{2}\right) \right)  \nonumber \\
&\equiv &f_{{\rm n}}\left( \varepsilon \right) \left( 1-f_{{\rm n}}\left(
\varepsilon -\omega ^{\prime }\right) \right) \left( 1-f_{l}\left( \mu
_{l}+\omega ^{\prime }-\omega _{1}\right) \right) ,  \label{block}
\end{eqnarray}%
where $\omega ^{\prime }\sim T$.

Furthermore, since the antineutrino energy is $\omega _{1}\sim T$, and the
antineutrino momentum $\left| {\bf k}_{1}\right| \sim T$ is much smaller
than the momenta of other particles, we can neglect the neutrino
contributions in the energy-momentum conserving delta-functions%
\begin{eqnarray}
&&\delta \left( \varepsilon +U_{{\rm n}}-\varepsilon ^{\prime }-U_{{\rm p}%
}-\omega _{1}-\omega _{2}\right) \delta \left( {\bf p}-{\bf p}^{\prime }-%
{\bf k}_{1}-{\bf k}_{2}\right)  \nonumber \\
&\simeq &\delta \left( \varepsilon +U_{{\rm n}}-\varepsilon ^{\prime }-U_{%
{\rm p}}-\omega _{2}\right) \delta \left( {\bf p}-{\bf p}^{\prime }-{\bf k}%
_{2}\right)  \label{delE}
\end{eqnarray}%
and perform integral over $d^{3}p^{\prime }$ to obtain ${\bf p}^{\prime }=%
{\bf p}-{\bf k}_{2}$ in the next integrals.

The energy exchange in the matter goes naturally on the temperature scale $%
\sim T$, which is small compared to typical kinetic energies of degenerate
particles. Therefore, in all smooth functions under the integral (\ref{gold}%
), the momenta of in-medium fermions can be fixed at their values at Fermi
surfaces, which we denote as $p_{{\rm n}}$, $p_{{\rm p}}$ for the nucleons
and $p_{l}$ for leptons respectively.

The energy of the final lepton is close to its Fermi energy $\mu _{l}=\mu _{%
{\rm n}}-\mu _{{\rm p}}$, and the chemical potentials of nucleons can be
approximated by their individual Fermi energies $\mu _{{\rm n}}=\varepsilon
_{{\rm n}}+U_{{\rm n}}$, $\mu _{{\rm p}}=\varepsilon _{{\rm p}}+U_{{\rm p}}$%
. This allows us to transform the energy-conserving $\delta $-function as 
\begin{eqnarray}
&&\delta (\varepsilon _{{\rm n}}-\sqrt{p_{{\rm n}}^{2}+p_{l}^{2}-2p_{{\rm n}%
}p_{l}\cos \theta _{l}+M^{\ast \,2}}+U_{{\rm n}}-U_{{\rm p}}-\mu _{l}) 
\nonumber \\
&=&\frac{\varepsilon _{{\rm p}}}{p_{{\rm n}}p_{l}}\delta \left( \cos \theta
_{l}-\frac{1}{2p_{{\rm n}}p_{l}}\left( p_{{\rm n}}^{2}-p_{{\rm p}%
}^{2}+p_{l}^{2}\right) \right) ,  \label{delfun}
\end{eqnarray}%
where $\theta _{l}$ is the angle between the momentum ${\bf p}_{{\rm n}}$ of
the initial neutron and the momentum ${\bf p}_{l}$ of the final lepton.
Notice, when the baryon and lepton momenta are at their individual Fermi
surfaces, the $\delta $- function (\ref{delfun}) does not vanish only if $p_{%
{\rm p}}+p_{l}>p_{{\rm n}}$.

Further we use the particular frame with $Z$-axis directed along the neutron
momentum ${\bf p}_{{\rm n}}$. Then%
\begin{eqnarray}
P_{1} &=&(0,\,0,\,p_{{\rm n}},\,\varepsilon _{{\rm n}})  \nonumber \\
k_{1} &=&\omega _{1}\left( \sin \theta _{\nu },\,0,\,\cos \theta _{\nu
},\,1\right)  \nonumber \\
k_{2} &=&\left( p_{l}\sin \theta _{l}\cos \varphi _{l},\,p_{l}\sin \theta
_{l}\sin \varphi _{l},\,p_{l}\cos \theta _{\nu },\,\mu _{l}\right)
\label{Pk}
\end{eqnarray}%
The energy-momentum of the final proton is defined by conservation laws:%
\begin{equation}
P_{2}=\left( -p_{l}\sin \theta _{l}\cos \varphi _{l},\,-p_{l}\sin \theta
_{l}\sin \varphi _{l},\,p_{{\rm n}}-p_{l}\cos \theta _{\nu },\,\varepsilon _{%
{\rm p}}\right)  \label{P2}
\end{equation}%
Insertion of (\ref{Pk}) and (\ref{P2}) in the square of the matrix element (%
\ref{MatrEl}) yields a rather cumbersome expression, which, however, is
readily integrable over solid angles of the particles.

Since we focus on the actually important case of degenerate nucleons and
leptons, we may consider the neutrino energy losses to the lowest accuracy
in $T/\mu _{l}$. Then the remaining integration reduces to the factor 
\begin{eqnarray}
&&\int d\omega _{1}\omega _{1}^{3}d\omega ^{\prime }d\varepsilon \,f_{{\rm n}%
}\left( \varepsilon \right) \left( 1-f_{{\rm n}}\left( \varepsilon -\omega
^{\prime }\right) \right) \left( 1-f_{l}\left( \mu _{l}+\omega ^{\prime
}-\omega _{1}\right) \right)   \nonumber \\
&\simeq &\int_{-\infty }^{\infty }d\omega ^{\prime }\frac{\omega ^{\prime }}{%
\exp \omega ^{\prime }/T-1}\int_{0}^{\infty }d\omega _{1}\frac{\omega
_{1}^{3}}{1+\exp \left( \omega _{1}-\omega ^{\prime }\right) /T}=\frac{457}{%
5040}\pi ^{6}T^{6}.  \label{int}
\end{eqnarray}%
Finally the neutrino emissivity is found to be of the form\footnote{%
See footnote to Eq. (\ref{MatrEl})}:%
\begin{eqnarray}
Q &=&\,\frac{457\pi }{10\,080}G_{F}^{2}C^{2}T^{6}\Theta \left( p_{l}+p_{{\rm %
p}}-p_{{\rm n}}\right) \left\{ \left( C_{A}^{2}-C_{V}^{2}\right) M^{\ast
2}\mu _{l}\right.   \nonumber \\
&&+\frac{1}{2}\left( C_{V}^{2}+C_{A}^{2}\right) \left[ 4\varepsilon _{{\rm n}%
}\varepsilon _{{\rm p}}\mu _{l}-\left( \varepsilon _{{\rm n}}-\varepsilon _{%
{\rm p}}\right) \left( \left( \varepsilon _{{\rm n}}+\varepsilon _{{\rm p}%
}\right) ^{2}-p_{l}^{2}\right) \right]   \nonumber \\
&&+C_{V}C_{M}\frac{M^{\ast }}{2M}\left[ 2\left( \varepsilon _{{\rm n}%
}-\varepsilon _{{\rm p}}\right) p_{l}^{2}-\left( 3\left( \varepsilon _{{\rm n%
}}-\varepsilon _{{\rm p}}\right) ^{2}-p_{l}^{2}\right) \mu _{l}\right]  
\nonumber \\
&&+C_{A}\left( C_{V}+\frac{M^{\ast }}{M}C_{M}\right) \left( \varepsilon _{%
{\rm n}}+\varepsilon _{{\rm p}}\right) \left( p_{l}^{2}-\left( \varepsilon _{%
{\rm n}}-\varepsilon _{{\rm p}}\right) ^{2}\right)   \nonumber \\
&&+C_{M}^{2}\frac{1}{16M^{2}}\left[ 8M^{\ast 2}\left( \varepsilon _{{\rm n}%
}-\varepsilon _{{\rm p}}\right) \left( p_{l}^{2}-\left( \varepsilon _{{\rm n}%
}-\varepsilon _{{\rm p}}\right) \mu _{l}\right) \right.   \nonumber \\
&&+\left( p_{l}^{2}-\left( \varepsilon _{{\rm n}}-\varepsilon _{{\rm p}%
}\right) ^{2}\right) \left( 2\varepsilon _{{\rm n}}^{2}+2\varepsilon _{{\rm p%
}}^{2}-p_{l}^{2}\right) \mu _{l}  \nonumber \\
&&\left. -\left( p_{l}^{2}-\left( \varepsilon _{{\rm n}}-\varepsilon _{{\rm p%
}}\right) ^{2}\right) \left( \varepsilon _{{\rm n}}+\varepsilon _{{\rm p}%
}\right) ^{2}\left( 2\varepsilon _{{\rm n}}-2\varepsilon _{{\rm p}}-\mu
_{l}\right) \right]   \nonumber \\
&&\left. -C_{A}^{2}M^{\ast 2}\Phi \left( 1+m_{\pi }^{2}\Phi \right) \left[
\mu _{l}\left( \left( \varepsilon _{{\rm n}}-\varepsilon _{{\rm p}}\right)
^{2}+p_{l}^{2}\right) -2\left( \varepsilon _{{\rm n}}-\varepsilon _{{\rm p}%
}\right) p_{l}^{2}\right] \right\}   \label{QMFA}
\end{eqnarray}%
with $\Theta \left( x\right) =1$ if $x\geq 0$ and zero otherwise. In the
above, the last term, with $\allowbreak $ 
\begin{equation}
\Phi =\frac{1}{m_{\pi }^{2}+p_{l}^{2}-\left( \varepsilon _{{\rm n}%
}-\varepsilon _{{\rm p}}\right) ^{2}},  \label{FI}
\end{equation}%
represents the contribution of the pseudoscalar interaction. The
''triangle'' condition $p_{{\rm p}}+p_{l}>p_{{\rm n}}$, required by the
step-function, is necessary for conservation of the total momentum in the
reaction and exhibits the threshold dependence on the proton concentration.

Neutrino energy losses caused by the direct Urca on nucleons depend
essentially on the composition of beta-stable nuclear matter. Therefore, in
order to estimate the relativistic effects, we consider the model of nuclear
matter, which besides nucleons includes $\Sigma $ and $\Lambda $ hyperons %
\cite{PBPELK97}. The parameters of the model are chosen as suggested in Ref. %
\cite{GM} to reproduce the nuclear matter equilibrium density, the binding
energy per nucleon, the symmetry energy, the compression modulus, and the
nucleon effective mass at saturation density $n_{0}=0.16$ $fm^{-3}$. The
composition of neutrino-free matter in beta equilibrium among nucleons,
hyperons, electrons and muons is shown on the left panel of Fig. 1 versus
the baryon number density $n_{b}$, in units of $n_{0}$.

On the right panel of Fig. 1, by a solid curve we show the relativistic
neutrino emissivity of reactions $n\rightarrow p+e^{-}+\bar{\nu}_{e}$, $\ \
\ p+e^{-}\rightarrow n+\nu _{e}$, as given by Eq. (\ref{QMFA}). Appearance
of hyperons in the system suppresses the nucleon fractions and lepton
abundance. Therefore at densities, where the number of hyperons is
comparable with the number of protons, the relativistic emissivity reaches
the maximum and then has a tendency to decrease. To inspect the contribution
of weak magnetism and the pseudoscalar interaction we demonstrate two
additional graphs. The long-dashed curve demonstrates the energy losses
obtained from Eq. (\ref{QMFA}) by formal setting $\Phi =0$. This eliminates
the pseudoscalar contribution. The dot-dashed curve is obtained by formal
replacing $\Phi =0$ and $C_{M}=0$, which eliminates both the weak magnetism
and pseudoscalar contributions. A comparison of these curves demonstrates
the weak magnetism effects. The contribution of the pseudoscalar interaction
can be observed by comparing the total neutrino energy losses (solid curve)
with the long-dashed curve, which is calculated without this contribution.
We see that, due to weak magnetism effects, relativistic emissivities
increase by approximately 40-50\%, while the pseudoscalar interaction only
slightly suppresses the energy losses, approximately by 5\%.

\begin{acknowledgments}
This work was partially supported by the Russian Foundation for Fundamental
Research Grant 00-02-16271
\end{acknowledgments}

\newpage
\vskip 0.3cm \psfig{file=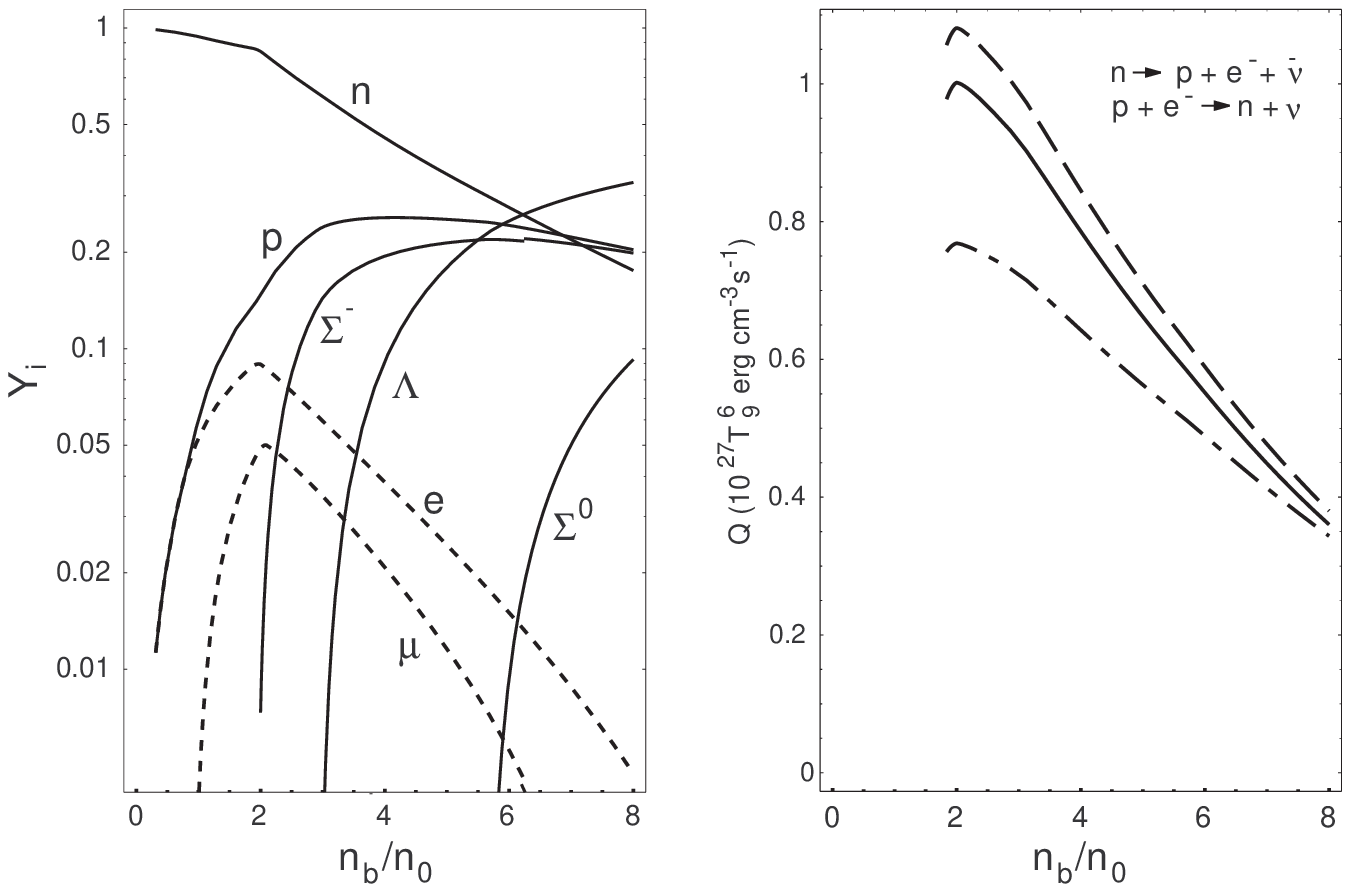} Fig. 1. The left panel shows individual
concentrations for matter in beta equilibrium among nucleons, hyperons,
electrons and muons as a function of the density ratio $n_{b}/n_{0}$. The
right panel represents the neutrino emissivity of the direct Urca processes
among nucleons and electrons for the matter composition represented on the
left panel. The curves begin at the threshold density. The solid curve
represents the total relativistic emissivity, as given by Eq. (\ref{QMFA}).
The dot-dashed curve shows the relativistic emissivity without contributions
of weak magnetism and pseudoscalar interaction, and the long-dashed curve is
the emissivity without the pseudoscalar contribution. All the emissivities
are given in units $10^{27}T_{9}^{6}$ $erg\,cm^{-3}s^{-1}$, where the
temperature $T_{9}=T/10^{9}\,K$. \vskip 0.3cm

\end{document}